\documentclass{cai}
\usepackage{cite}
\usepackage{amsmath,amssymb,amsfonts}
\usepackage{algorithm} 
\usepackage{algorithmic}
\usepackage{graphicx}
\usepackage{textcomp}
\usepackage{array}

 \volume{xx}
 \yyear{2021}
 \page{1001}

\begin{document}
\label{firstpage}

\title[Multi-stream CNN with Frequency Selection for Robust Speaker Verification]
{Multi-stream Convolutional Neural Network with Frequency Selection for Robust Speaker Verification}

\author[W. Yao]
{Wei \surname{Yao}*}
\affiliation{Key Laboratory of Technology in Rural Water Management of Zhejiang Province\\
College of Electric Engineering\\
Zhejiang University of Water Resources and Electric Power\\
Hangzhou, China}
\email{yaowei@zjweu.edu.cn}

\author[S. Chen]
{Shen \surname{Chen}*}
\affiliation{Delta Electronics (China)\\
Hangzhou, China}
\email{chenshen@next-aiot.cn}

\author[J. Cui]
{Jiamin \surname{Cui}}
\affiliation{Key Laboratory of Technology in Rural Water Management of Zhejiang Province\\
College of Electric Engineering\\
Zhejiang University of Water Resources and Electric Power\\
Hangzhou, China}
\email{cuijm@zjweu.edu.cn}

\author[Y. Lou]
{Yaolin \surname{Lou}}
\affiliation{Key Laboratory of Technology in Rural Water Management of Zhejiang Province\\
College of Electric Engineering\\
Zhejiang University of Water Resources and Electric Power\\
Hangzhou, China}
\email{louyl@zjweu.edu.cn}

%
%

\noreceived{} \nocommunicated{}

\maketitle

\begin{abstract}
Speaker verification aims to verify whether an input speech corresponds to the claimed speaker, 
and conventionally, this kind of system is deployed based on single-stream scenario, wherein 
the feature extractor operates in full frequency range.
In this paper, we hypothesize that machine can learn enough knowledge to do classification task 
when listening to partial frequency range instead of full frequency range, which is so called 
frequency selection technique, and further propose 
a novel framework of multi-stream Convolutional Neural Network (CNN) 
with this technique for speaker verification tasks.
The proposed framework accommodates diverse temporal embeddings generated from 
multiple streams to enhance the robustness of acoustic modeling. 
For the diversity of temporal embeddings, we consider feature augmentation with frequency selection, 
which is to manually segment the full-band of frequency into several sub-bands,
and the feature extractor of each stream can select which sub-bands to use as target frequency domain.
Different from conventional single-stream solution wherein each utterance would only be processed for one time,
in this framework, there are multiple streams processing it in parallel. 
The input utterance for each stream is pre-processed by a frequency selector within specified frequency range,
and post-processed by mean normalization. 
The normalized temporal embeddings of each stream will flow into a pooling layer to generate fused embeddings.
We conduct extensive experiments on VoxCeleb dataset, 
and the experimental results demonstrate that multi-stream CNN significantly outperforms 
single-stream baseline with 20.53 \% of relative improvement in minimum Decision Cost Function (minDCF).
\end{abstract}


\begin{mathclass}
97R40
\end{mathclass}

\section{Introduction}
Deep learning has achieved outstanding success in various speech-oriented tasks, 
such as auto speech recognition \cite{b1,b2}, speaker recognition \cite{b3}--\cite{b6} 
and speaker diarization \cite{b7,b8}, etc. The deep learning paradigm is addressed to extract highly
abstracted representations by means of well-designed neural networks based on the feed-in data.
Most commonly, there are three scenarios to train neural network in deep learning, 
which are supervised learning \cite{b9}, semi-supervised learning \cite{b10} and 
unsupervised learning (or, more precisely, self-supervised learning) \cite{b11}, respectively. 
In addition, supervised learning with abundant labeled data is the most widely used scenario \cite{b12}, 
which is also the scenario used in this paper.

Speaker recognition is the filed of recognizing speaker identities based on their voices. 
In general, it can be clarified into either 1) speaker verification or 2) speaker identification. 
Speaker verification aims to answer the question ``is somebody speaking?'' with single utterance  or 
``are they from the same speaker?'' with pairwise utterances. Speaker identification is used to answer 
the question ``who is speaking?'' among a set of enrolled speakers. 
Speaker verification is one case of biometric authentication, 
where user provides their biometric characteristics in form of voiceprint as passwords.
The greatest challenge of speaker verification task is the effective usage of datasets 
obtained from the real world under noisy and unconstrained conditions \cite{b13}.
In this paper, we aim to address this challenge and propose a new framework to extract robust speaker embeddings.

\subsection{Frequency Selection}
Normally, the features used for training and testing are extracted in full frequency band,
and they are usually low dimensional.
As features play an important role in speaker verification system, if we use just partial frequency range
instead of full frequency range, will the system perform equally well?
Follow by this assumption, we first segment full-band into several sub-bands by using frequency selector,
e.g., low frequency sub-band and high frequency sub-band, 
and use these features to train several single-stream system respectively, eager to witness
the impact of frequency domain on system performance.
The feature extractor can select which sub-bands to use as target frequency domain to generate frame-level features,
and we call this idea as frequency selection.

\subsection{Resolution and Problem Statement}
We hypothesize that machine could dramatically benefit from our proposed frequency selection technique. 
On the other hand, Convolutional Neural Network (CNN) is one kind of widely used neural networks in image recognition. 
More recently, CNN is introduced to speaker recognition and achieves competitive results \cite{b5,b6}, compared with 
the most famous Time-Delay Neural Network (TDNN) and its variations \cite{b14}. Despite demonstrating encouraging outcomes, 
CNN for speaker verification is remaining an open topic, which requires more efforts to achieve breakthrough.
By investigating various ways of doing so, we bridge frequency selection and CNN,
and propose a new framework of multi-stream CNN. 

However, this approach may prompt some questions: 
Why to use multi-stream if single-stream can offer us a high enough accuracy? 
Can machine learn enough knowledge to handle classification task by only listening to partial frequency range?
Demonstrated by our experimental results, these are part of questions we are going to delve into and 
find out answer in this paper.

\subsection{Contributions}
Most of the speaker verification systems are deployed based on single-stream within full frequency range.
To the best of our knowledge, this paper is the first to investigate the impact of frequency domain
and to improve system performance by means of frequency selection. 
Our contributions are as follows:

(1) We explore the performance of neural network by feeding in ``partial'' features extracted from
speech within sub-bands of frequency instead of conventionally used full-band. 
And we find that machine can perform equivalently well in some sub-bands, which are also  
beneficial to improve the performance of multi-stream system.

(2) We propose the idea of frequency selection, 
and a novel framework of multi-stream CNN based on it for speaker verification.

(3) We make our work open source, 
and it is available to download at 

\underline{https://github.com/ShaneRun/multistream-CNN}.

\subsection{Structure of This Paper}
This paper is organized as below. We review on the related work in Section~\ref{sec:relatedwork}. 
Section~\ref{sec:proposedmethod} describes our proposed method in detail. 
Section~\ref{sec:experiments} presents experiments for pairwise verification, which consists of 
dataset, training and results. We also make comprehensive comparison in this section in order to demonstrate
the efficacy and understand the influence of frequency selection.
Section~\ref{sec:futurework} contains discussion and future work.
Section~\ref{sec:conclusion} is the conclusion of our work.

\section{Related Work}
\label{sec:relatedwork}
\begin{figure}
    \begin{center}    
    \includegraphics[width=1.0\textwidth]{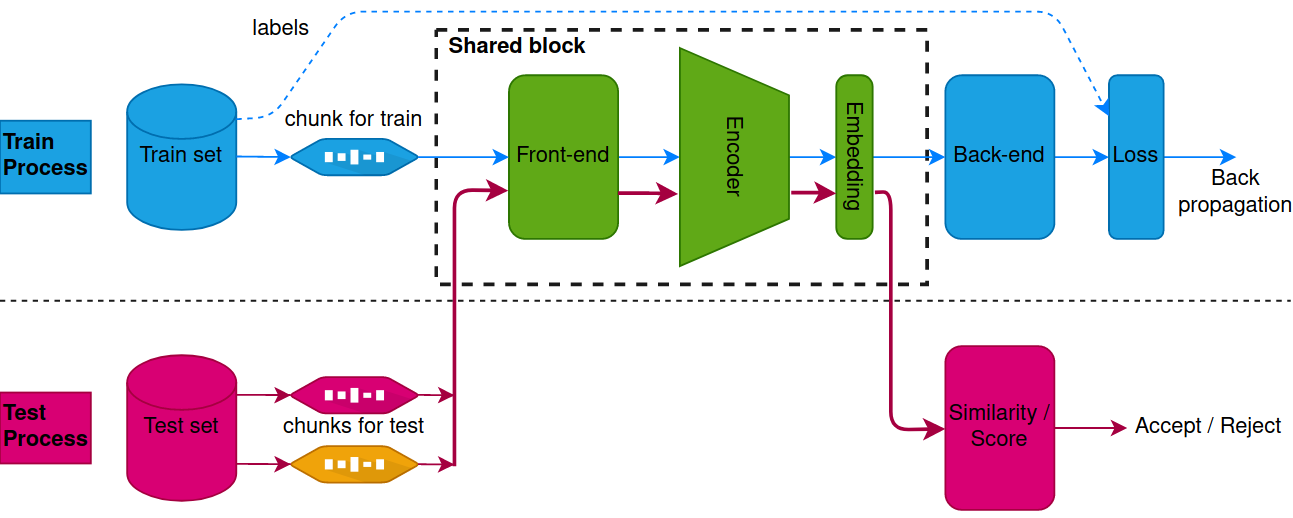}    
    \end{center}    
    \caption{A schematic illustration of a speaker verification system consisting of train process and test process.
    The train process is to tune the parameters of encoder in order to minimize the loss using back propagation. 
    The test process is to make a decision on accept or reject under the tested pair of utterances from the test set.}    
    \label{fig1}    
    \end{figure}

Most of works done so far on speaker verification are based on single-stream framework as illustrated in Figure~\ref{fig1}.
In general, it is comprised of train process (including validation) and test process,
and can be classified into several modules, including front-end, encoder, back-end, loss and similarity/score.
The train process is to tune network parameters of encoder using abundant labeled data.
After training, the back-end and loss module are not used any longer, whereas the shared block 
(including front-end and encoder), which is enclosed by imaginary line, will be inherited by the test process.
The test process is to make a decision on whether or not the utterance pair is from the same speaker.

The front-end module, or rather feature extractor, is used to extract quality frame-level features for subsequent
signal processing by converting acoustic waveform into a relatively lower dimensional representations, 
such as the well known Mel Filter Bank Energies (MFBE) \cite{b15,b16} and Mel Frequency Cepstral Coefficients (MFCC) \cite{b17,b18}.
The encoder is used to extract unique speaker embeddings in utterance-level through deep neural network,
which is known as identity vector, e.g., ``i-vector'' \cite{b19}, ``x-vector'' \cite{b20}, and ``r-vector'' \cite{b21}.
The back-end module is for post-processing of speaker embeddings in order to enlarge inter-class distance and
reduce intra-class distance. The most popular approaches for back-end processing contain Gaussian back-end model \cite{b22},
Probabilistic Linear Discriminant Analysis (PLDA) \cite{b23}, and neural-based model \cite{b24}.
The loss module plays an important role in training because it is actually defining the goal function, and
this module is no longer functional after training. In \cite{b6}, extensive evaluations of the most popular loss function, 
including softmax loss, angular softmax loss, triplet loss and angular prototypical loss, etc., are presented,
and it is also demonstrated that metric learning objectives outperform classification-based losses.
The similarity module generates probability or score based on the embeddings pair by adopting Euclidean distance
or Cosine similarity, and then makes a decision on accept or reject.

Over the years, multi-stream approach has attracted a lot of attentions in deep learning field, 
such as Computer Vision (CV) \cite{b25}--\cite{b27} and Automatic Speech Recognition (ASR) \cite{b28}--\cite{b38}. 
\cite{b25} and \cite{b26} both propose a multi-stream CNN architecture to recognize human actions and gestures. 
\cite{b25} is implemented by combining novel human-related streams containing one appearance and one motion stream with
the traditional streams. Whereas \cite{b26} decomposes the original image into several equal-sized streams and learn 
representations by a CNN for each stream. Then the features learned from all streams are fused into a unified feature map,
which is subsequently fed into a neural network to recognize gestures.
In \cite{b27}, a multi-stream framework, which is comprised of motion stream, spatial stream and structural stream,
is designed for unmanned aerial vehicles video aesthetic quality assessment.

In \cite{b28}, a novel effort to estimate word error rate uses a multi-stream end-to-end architecture
based on a combination of four independent streams which deal with decoder, acoustics, textual and 
phonotactics features in parallel.
\cite{b29,b30,b31,b32} all employ several parallel streams by using audio-visual strategy. 
The reason behind these approaches is to address the problem of speech recognition
by leveraging visual information to improve the performance of ASR. 
Nevertheless, \cite{b33}--\cite{b38} aim to capture diverse information from audio only for end-to-end ASR. 
For instance, \cite{b33} presents a framework based on joint attention with multiple audio streams in parallel.
More practically, in order to address of problem of massive computation and memory requirements during training 
resulting from increasing number of streams, \cite{b34} introduces a two-stage training scenario for end-to-end ASR,
where training of feature extractor and attention fusion module are processed in separated stages.

It can be perceived that multi-stream framework is successfully used in ASR inspired by observing multiple
streams in parallel. However, there are still limited research on speaker verification tasks.
To the best our knowledge, \cite{b35} is the most related works that has been done so far on speaker verification task,
but in this work the multiple streams are generated after feature extraction based on the trained intelligibility likelihood model.
In addition, it only uses multi-stream in the test process.
On the contrary, in our work, we first form streams by selecting the original speech signal in frequency domain,
and then design our framework to make use of multiple speech streams in both train and test process.

Materials and Methods should be described with sufficient details to allow others to replicate and build on published results. Please note that publication of your manuscript implicates that you must make all materials, data, computer code, and protocols associated with the publication available to readers. Please disclose at the submission stage any restrictions on the availability of materials or information. New methods and protocols should be described in detail while well-established methods can be briefly described and appropriately cited.

Research manuscripts reporting large datasets that are deposited in a publicly avail-able database should specify where the data have been deposited and provide the relevant accession numbers. If the accession numbers have not yet been obtained at the time of submission, please state that they will be provided during review. They must be provided prior to publication.

Interventionary studies involving animals or humans, and other studies require ethical approval must list the authority that provided approval and the corresponding ethical approval code.

\section{Proposed Method}
\label{sec:proposedmethod}
In this section, we propose a multi-stream framework using three streams processing in parallel
for speaker verification task. We present design details of all modules containing frequency selection,
acoustic feature, ResNet-34 based encoder, loss function and similarity.

\subsection{Diagram of Framework}
\begin{figure}
    \begin{center}    
    \includegraphics[width=1.0\textwidth]{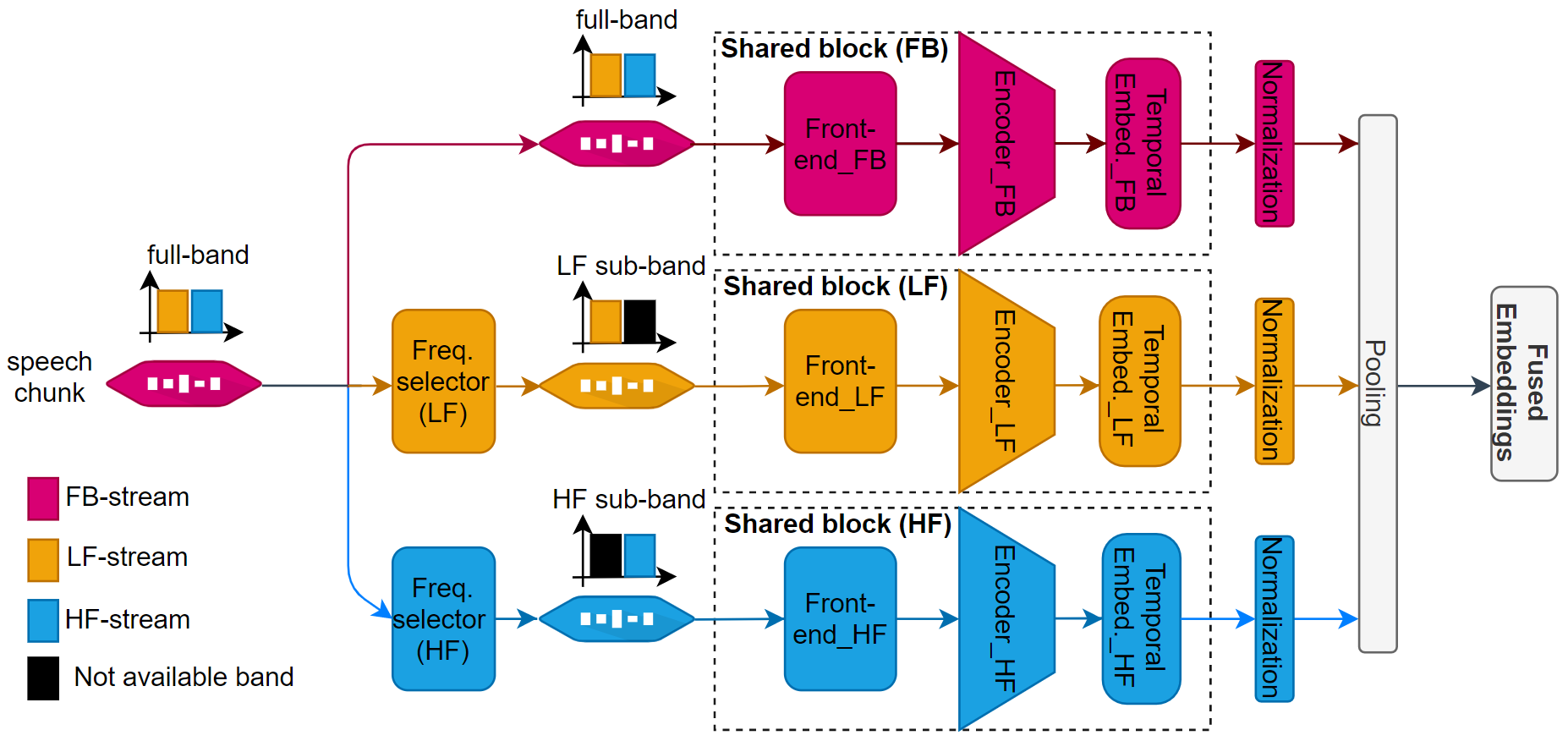}    
    \end{center}    
    \caption{Proposed framework of multi-stream CNN.
    Shared block (FB), Shared block (LF) and Shared block (HF) have the same network structure.}    
    \label{fig2}    
    \end{figure}

The proposed framework of multi-stream CNN is illustrated in Figure~\ref{fig2}.
Frequency full-band (FB) is segmented into two sub-bands after frequency selector, including
Low Frequency (LF) sub-band and High Frequency (HF) sub-band.
The same speech signal is processed by different streams in parallel, 
namely FB-stream, LF-stream, and HF-stream, respectively.
In our proposed framework, 
FB-stream serves as a general encoder because it listens to full frequency band, 
whereas LF-stream and HF-stream encoders serve as specialized encoder because they focus on 
listening to LF-band and HF-band of input speech.
Even though the overall performance of LF-stream or HF-stream are relatively poorer than FB-stream
(this is also demonstrated by our experiments in the next section),
they can still contribute to performance of our proposed multi-stream system,
mainly because they are more good at extracting partial characteristics, so that 
the fused speaker embeddings can be more robust. 

The shared block of each stream has the same structure but different in parameters.
After temporal embeddings are extracted by all streams, they are then mean-normalized, 
and subsequently fed into a pooling layer to generate fused or final embeddings.

Similar to conventional acoustic modeling, each stream encodes the acoustic features into highly abstracted
temporal embeddings as formulated in \eqref{eq1}:
\begin{equation}\vec{x}^{(s)}=Encoder^{(s)}(\vec{u}), s\in{\{1,2,3\}}\label{eq1}\end{equation}
where superscript $s\in{\{1,2,3\}}$ is denoted as index for each encoder of corresponding stream $s$ 
(with 1,2 and 3, for FB-stream, LF-stream and HF-stream, respectively), 
$\vec{u}$ is the input vector of chunk which is usually extracted with fixed length (2 - 4 seconds)
randomly from speech utterances (note that each stream use the same $\vec{u}$),
$\vec{x}$ is the output vector of encoder which is denoted as temporal embeddings in this article. 

Then the fused embeddings in terms of $\vec{x_f}$ can be formulated as:
\begin{equation}\vec{x_f}=\sum\limits_{s=1}^{3}{k_t^{(s)}{\vec{x}^{(s)}}}\label{eq2}\end{equation}
where $k_t^{(s)}$ is the fusion weight for each temporal embeddings, 
and the definition of $s$ is the same as \eqref{eq1}.

\subsection{Frequency Selection}
Normally, the acoustic features are obtained with full-band of frequency.
In this article, we want to witness machine's capability to do speaker verification with different sub-bands of frequency.
We use frequency selector on top of each stream, and by doing so, 
the upper and lower limit of different sub-bands can be configurable.

\begin{figure}
    \begin{center}    
    \includegraphics[width=0.7\textwidth]{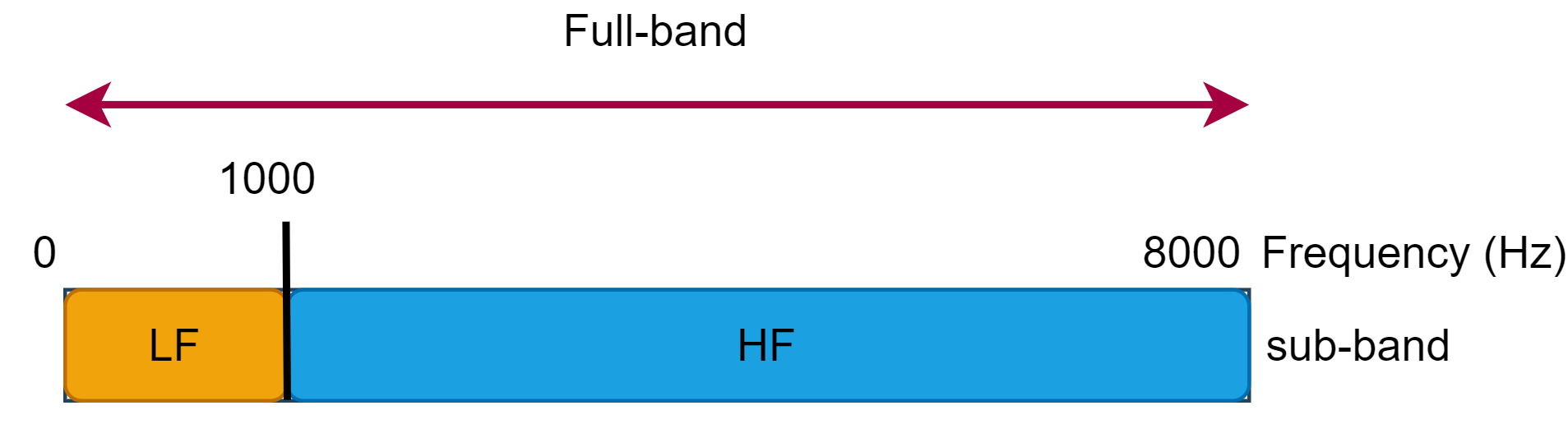}  
    \end{center}      
    \caption{Frequency Segmentation}    
    \label{fig3}    
    \end{figure}

More specifically, the boundary between LF and HF sub-band is initially set to 1000 Hz.
The reason behind this setting is that the fundamental frequency of the complex speech tone 
which is also known as the pitch or $f_0$, lies above 500 Hz even though it differs from person to person,
whereas this range might extend approximately to 1000 Hz in some cases \cite{b39}. 
The overview of our frequency segmentation is shown in Figure~\ref{fig3}.

We implement frequency selection in PyTorch \cite{b40} by using a class called ``torchaudio.transforms.MelSpectrogram'',
which is designed to create MelSpectrogram for a raw audio signal.
In practice, we can adjust minimum frequency ($f_{min}$, default value is 0) and 
maximum frequency ($f_{max}$, default value is half of the sampling rate) in our training script so that
frequency selection will be employed.

\subsection{Acoustic Feature}
The most common forms of acoustic feature are Mel Filter Bank Energies (MFBE) and 
Mel Frequency Cepstral Coefficients (MFCC). 
The procedures for computing of MFBE and MFCC features are similar, 
where in both cases the speech signal is first pre-processed by a pre-emphasis filter;
then it is segmented into frames with overlapping and all frames are applied with 
a window function (normally hamming window) in order to reduce spectrum leakage; 
afterwards, each frame goes through a Fourier transform to calculate filter bank energies
(or power spectrum).
To obtain MFCC, a Discrete Cosine Transform (DCT) is applied to the filter bank energies
in order to retain the resulting coefficients, whereas the other coefficients will be discarded \cite{b41}.
The reason for discarding the other coefficients is that they represent fast changes 
in the filter bank coefficients and these fine details seldom contribute to system performance.

Despite the huge contribution of MFCC to speech-related tasks,
there is something wrong with it as mentioned above: it requires a extra step called DCT 
in order to decorrelate coefficients of filter bank energies.
Since DCT is an additional linear transform, some information in highly non-linear speech signals 
will be discarded undesirably. 
For this reason, MFBE is becoming increasingly popular due to rapid growth of end-to-end techniques using deep learning,
because it can provide more information than MFCC for neural networks to delve into \cite{b15,b16,b42}.

We use 40-dimensional MFBE \cite{b43} as acoustic feature, with a hamming window of 25 ms width and 10 ms step.
The chunk length is extracted randomly from each utterance, 
and is fixed as 2 seconds and 4 seconds for training and testing, respectively.

\subsection{ResNet-34 Based Encoder}
ResNet, short for Residual Network, is a form of CNN introduced in \cite{b44}, 
and achieves extreme success in the field of image recognition.
The basic idea of ResNet is to alleviate the problem of gradient vanish or explosion 
in training very deep neural networks by using residual blocks with skip connections. 
The skip connections behave as shortcut paths for gradient to flow through alternately, 
and allow the higher layer to learn the identity functions directly
so that it can perform not worse than the lower layer.

For the last few years, some works are done to introduce ResNet to the field of speaker recognition,
and achieves encouraging results \cite{b5,b6}, compared with TDNN and its variations \cite{b14}.
We assume speaker recognition is somewhat the same as image recognition, and therefore
ResNet can also become very popular in this field.

We use ResNet-34 \cite{b44} with 34 hidden layers as network structure of encoder as shown in Table~\ref{tab1}.
The total frames is 200 for each chunk, therefore, the size of input feature is $ 40\times 200$.
The network consists of 34 convolutional layers with batch normalization and 
Rectified Linear Units (ReLU) activation function applying to each of them,
and these layers can be grouped into Conv1, Res1, Res2, Res3, Res4 and Flatten, respectively.
The output of encoder is 512-dimensional speaker embeddings.

\begin{table}
    \caption{Network Structure of ResNet-34 Based Encoder. 
    ASP: Attentive Statistics Pooling.}
    \label{tab1}
    \centering
    \setlength{\tabcolsep}{3pt}
    \begin{tabular}{|m{80pt}|m{100pt}|m{80pt}|m{80pt}|}
    \hline
    Layer group& 
    Kernel size& 
    Stride$^{\mathrm{a}}$&
    Output size\\
    \hline
    Input& 
    --&
    --& 
    $40 \times 200 \times 1$\\
    Conv1& 
    $3 \times 3\times 16$&
    $1 \times 1$& 
    $40 \times 200 \times 16$\\
    Res1& 
    $$\left[\begin{array}{lll}3 \times 3 \times 16\\3 \times 3 \times 16\end{array}\right] \times 3$$&
    $1 \times 1$& 
    $40 \times 200 \times 16$\\   
    Res2& 
    $$\left[\begin{array}{lll}3 \times 3 \times 32\\3 \times 3 \times 32\end{array}\right] \times 4$$&
    $2 \times 2$& 
    $20 \times 100 \times 32$\\
    Res3& 
    $$\left[\begin{array}{lll}3 \times 3 \times 64\\3 \times 3 \times 64\end{array}\right] \times 6$$&
    $2 \times 2$& 
    $10 \times 50 \times 64$\\
    Res4& 
    $$\left[\begin{array}{lll}3 \times 3 \times 128\\3 \times 3 \times 128\end{array}\right] \times 3$$&
    $2 \times 2$& 
    $5 \times 25 \times 128$\\
    Flatten& 
    --&
    --& 
    $5 \times 2048$\\
    ASP& 
    --&
    --& 
    4096\\
    Linear& 
    512&
    --& 
    512\\
    \hline
    \multicolumn{4}{p{350pt}}{$^{\mathrm{a}}$For stride that is not 1, it is only performed on the
    top layer of each residual block for down sampling, whereas the stride inside residual block 
    is always 1.}
    \end{tabular}
    \end{table}

\subsection{Loss Function}
Loss function plays an important role in training neural networks because it estimates
the error for the current state, which is then used to update the weights of the model 
through gradient descent and back propagation. By doing it repeatedly for massive times,
the overall loss of the model tends to be minimized which is usually expected to be 
the global minimum.

\emph{Softmax Loss.} The Softmax loss is one typical form of loss for multi-class classification tasks,
and it can accept many inputs and calculate probability for each one. 
The Softmax loss (in terms of $L_{sm}$) consists of a Softmax function followed by a cross-entropy loss,
which is formulated as \eqref{eq3}:
\begin{equation}L_{sm}=-\frac{1}{N}\sum\limits_{i=1}^{N}{
    \log\frac{
        e^{\vec{W}_{y_i}^T\vec{x_i}+\vec{b}_{y_i}}}{
            \sum_{j=1}^{C}{e^{\vec{W}_{j}^T\vec{x_i}+\vec{b}_{j}}}}}\label{eq3}\end{equation}

where $\vec{W}$ and $\vec{b}$ are the weight and bias vector of last layer of encoder, respectively.
$C$ is the total amount of classes (or speakers), and $N$ is the number of utterances of each mini-batch
from different speakers with embeddings $\vec{x_i}$ (extracted by encoder as defined in \eqref{eq1}) 
and its corresponding label $y_i$.

\emph{Angular Prototypical Loss.} Similar to the original prototypical loss,
the angular prototypical loss uses the same batch formation, whereas the similarity metric $S$
is changed from distance-based to cosine-based as shown in \eqref{eq4}:

\begin{equation}
    \vec{S}_{i,k} = \omega \times \cos(\vec{x}_{i,k}, \vec{c}_k) + b\label{eq4}\end{equation}
where $\omega$ and $b$ are learnable weight and bias,
$\vec{c}_k$ is the centroid (or prototype) as shown in \eqref{eq5}:

\begin{equation}
    \vec{c}_k = \frac{1}{M-1}\sum_{m=1}^{M-1}\vec{x}_{k,m}\label{eq5}\end{equation}
where $M$ is the utterances number for every speaker inside mini-batch.

The angular prototypical loss is then derived as:
\begin{equation}L_{ap}=-\frac{1}{N}\sum\limits_{i=1}^{N}{
    \log\frac{
        e^{\vec{S}_{i,i}}}{
            \sum_{k=1}^{N}{e^{\vec{S}_{i,k}}}}}\label{eq6}\end{equation}

\emph{Softmax + Angular Prototypical Loss.} The ultimate goal of designing loss function is to enlarge 
inter-class distance and meanwhile reduce inter-class distance.
To this end, we propose a combined form of Softmax loss $L_{sm}$ and Angular Prototypical loss $L_{ap}$ as 
fused loss function of our work, which is illustrated in \eqref{eq7}:

\begin{equation}
    L = L_{sm} + L_{ap}\label{eq7}\end{equation}

\subsection{Similarity}
Similarity is the scoring module used to compute the score based on the pair of speaker embeddings.
The score is used subsequently to make a decision on accept or reject.
Since open-set speaker recognition is essentially a metric learning problem, 
Euclidean distance \cite{b6} is more preferred as metric of similarity
than cosine similarity (measures the cosine of angle between two high-dimensional vectors) \cite{b45}.

The Euclidean distance between two speaker embeddings is formulated as:
\begin{equation}
    distance = \parallel(\vec{x_f}_1 - \vec{x_f}_2)\parallel_2 = 
    \sqrt{\sum_{d=1}^D{({x_f}_{1,d}-{x_f}_{2,d}})^2}\label{eq8}\end{equation}
where $\vec{x_f}_1$ and $\vec{x_f}_2$ are the fused embeddings of each speaker inside pair,
$D$ is the dimensions of embeddings which is designed to be 512 in this article.

\section{Experiments}
\label{sec:experiments}
In this section, we present the implementation details of our proposed framework.
Firstly, we introduce the VoxCeleb dataset used for training, validation and testing; 
Secondly, we present our training details and propose a practical training strategy inspired by \cite{b34};
Thirdly, we introduce the evaluation metrics used in our work;
And finally we describe our experimental results.

\subsection{Dataset}
Speaker verification faces many challenges in real situations related to ambient noise noise 
and short speech frame availability. 
Therefore, using a dataset generated from real world for experiments is more meaningful.
For this reason, We select VoxCeleb dataset for training and testing in our work.
The VoxCeleb dataset is obtained from real world, in which the speakers span a wide range of 
different ethnicities, accents, professions and ages. 
Speech of this dataset are shot in a large number of challenging auditory environments. 
Most crucially, all speech are degraded with real-world noise, consisting of background chatter, 
laughter, overlapping speech, room acoustics, 
and there is a range in the quality of recording equipment and channel noise \cite{b48}.
Moreover, the utterance length of VoxCeleb dataset distributes randomly from 4 seconds to 20 seconds.

VoxCeleb dataset is released by VGG group of Oxford university in two stages, 
as VoxCeleb1 \cite{b46} which contains 153,516 utterances (352 hours in total) from 1,251 celebrities, 
and VoxCeleb2 \cite{b47} which contains 1,128,246 utterances (2,442 hours in total) from 6,112 celebrities.
Moreover, there is challenge organized annually based on this dataset in order to witness how well 
current methods can recognize speakers from speech obtained ``in the wild'' since 2019 \cite{b13, b49}.

The datasets used in our work are listed in Table~\ref{tab2}. We use ``VoxCeleb2-Dev'' for training
without data augmentation, and cleaned version of ``VoxCeleb1-Test'' (also term as ``VoxCeleb1-O'') for validation and testing, 
which contains 37,720 testing pairs. 

\begin{table}
    \caption{Dataset for Training, Validation and Testing.}
    \label{tab2}
    \centering
    \setlength{\tabcolsep}{3pt}
    \begin{tabular}{|m{80pt}|m{80pt}|m{80pt}|m{80pt}|}
    \hline
    Stage& 
    Dataset& 
    \# of speakers&
    \# of utterances\\
    \hline
    Training& 
    VoxCeleb2-Dev$^{\mathrm{a}}$&
    5,994& 
    1,092,009\\
    Validation& 
    VoxCeleb1-Test$^{\mathrm{b}}$&
    40& 
    4,874\\    
    Testing& 
    VoxCeleb1-Test&
    40& 
    4,874\\   
    \hline
    \multicolumn{4}{p{320pt}}{$^{\mathrm{a}}$Development set of VoxCeleb2, which has no overlap
    with the identities in the VoxCeleb1. 
    $^{\mathrm{b}}$Test set of VoxCeleb1, cleaned version, the amounts of testing pairs is 37,720.}
    \end{tabular}
    \end{table}

\subsection{Training}
Our work is implemented in PyTorch framework with details showed in Table~\ref{tab3}.
All encoders are trained in a single GPU platform with 11 GB memory for maximum 100 epochs.
In order to reduce class imbalance, we apply random sampling with a maximum value of 100 utterances 
for each speaker in the training set. Additional, We use the largest batch size with 400
that fits on a GPU, and the training for one stream takes approximately three days.

\begin{table}
    \caption{Training Details Overview.}
    \label{tab3}
    \centering
    \setlength{\tabcolsep}{3pt}
    \begin{tabular}{|m{150pt}|m{200pt}|}
    \hline
    Item& 
    Value\\
    \hline
    Deep learning framework&
    PyTorch v1.5.1\\
    GPU&
    GeForce GTX 1080 TI (single)\\
    Optimizer&
    Adam\\
    Batch size&
    400\\
    Maximum epochs&
    100\\
    Initial learning rate&
    0.001\\
    Learning rate decay&
    0.95 per 10 epochs\\
    \hline
    \end{tabular}
    \end{table}

As the streams increased, conventional training approach wherein all encoders are trained in parallel,
is hard to be implemented due to massive computation and memory requirements.
In order to address this problem, we adopt a more practical training approach inspired by \cite{b34},
which consists of ``Stage 1: Sequential training of each stream'' and 
``Stage 2: Searching for optimal fusion weight''.

\emph{Stage 1: Sequential training of each stream.} In this stage, each stream will be trained
in sequential mode. Before training different streams, we only need to regulate the frequency range
which is designed in Figure~\ref{fig3}. In this way, training of large network becomes much more simple: 
to repeat the training of a relatively smaller network for multiple times. After sequential training,
the well-trained encoders will be deployed to the multi-stream system as shown in Figure~\ref{fig2}.

\emph{Stage 2: Searching for optimal fusion weight.} Based on the scores output of each stream,
we propose a simplified algorithm to address the problem of searching for optimal fusion weight.
As shown in Algorithm~\ref{alg1}, scores output of each stream is normalized in advance using 
t-Distributed Stochastic Neighbor Embedding (t-SNE) approach \cite{b50} and then used as input for this algorithm. 
$k_t^{(1)}$ is gradually reduced with $step$.
For every step of $k_t^{(1)}$, $k_t^{(2)}$ is also gradually reduced with $step$, 
and repeat calculating minDCF until $k_t^{(2)}$ is smaller than minimum weight $K_{min}$.
The return value will be used to update local optimum, which is defined as optimal weight under given $k_t^{(1)}$.
Local optimum will be used for updating global optimum for every step of $k_t^{(1)}$ repetitively until 
$k_t^{(1)}$ is smaller than minimum weight $K_{min}$.
In our experiment, $step$ and $K_{min}$ are set as 0.01 and 0, respectively.

\begin{algorithm}  
    \caption{Searching for Optimal Fusion Weight}  
    \label{alg1}  
    \begin{algorithmic} 
    \REQUIRE t-SNE normalized scores $\vec{scores}$ of each stream $s\in{\{1,2,3\}}$
    \ENSURE optimal fusion weight \{$k_t^{(1)}$, $k_t^{(2)}$, $k_t^{(3)}$\}
    \STATE $\vec{k_t} \leftarrow [1, 0, 0]$ 
    \REPEAT      
    \REPEAT
    \STATE $k_t^{(3)} \leftarrow (1.0 - k_t^{(1)} - k_t^{(2)})$
    \STATE calculate minDCF based on $\vec{scores}$ and $\vec{k_t}$
    \STATE update local optimum
    \STATE $k_t^{(2)} \leftarrow (k_t^{(1)} - step)$
    \UNTIL{$k_t^{(2)} \lneqq K_{min}$}
    \STATE update global optimum
    \STATE $k_t^{(1)} \leftarrow (k_t^{(1)} - step)$  
    \UNTIL{$k_t^{(1)} \lneqq K_{min}$}  
    \end{algorithmic}  
    \end{algorithm} 

\subsection{Evaluation Metrics}
\begin{figure}
    \begin{center}    
    \includegraphics[width=0.8\textwidth]{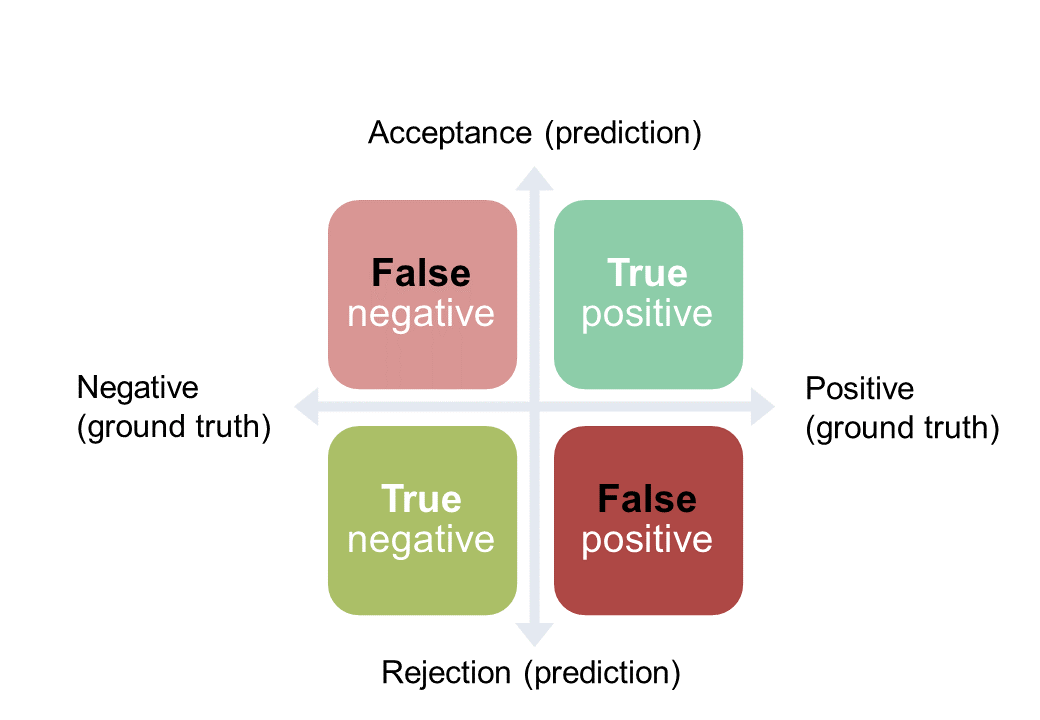}    
    \end{center}    
    \caption{Distribution Map of Decision Result.
    Ground truth is the label, and prediction is the score generated by the classifier}    
    \label{fig4}    
    \end{figure}

For speaker verification system with pairwise input, it is naturally a binary classifier, 
and the evaluation metrics are described in this section. 
The decision result distribution is illustrated in Figure~\ref{fig4}, which is comprised of 
True Positive (TP), True Negative (TN), False Positive (FP) and False Negative (FN).
The first term (True or False) is the result of prediction. 
If the prediction fits the ground truth, then the result is True, otherwise the result will be False. 
The second term (Positive or Negative) is the category of ground truth. If the utterance pair is 
from the same speaker, then the ground truth is Positive, otherwise it will be Negative.
Based on Figure~\ref{fig4}, we introduce two common kinds of evaluation metrics, which are 
Equal Error Rate (EER) and minimum Decision Cost Function (minDCF).

\emph{EER.} This is a widely used metric to determine the threshold value 
when false acceptance rate $E_{FA}$ and false rejection rate $E_{FR}$ are equal.

\begin{equation}
    E_{FA} = \frac{N_{FN}}{N_{FN}+N_{TP}}\label{eq9}\end{equation}

\begin{equation}
    E_{FR} = \frac{N_{FP}}{N_{FP}+N_{TN}}\label{eq10}\end{equation}

where $N_{FN}$, $N_{TP}$, $N_{FP}$ and $N_{TN}$ are the number of False Negative, True Positive, 
False Positive and True Negative, respectively.

\emph{minDCF.} National Institute of Standards and Technology (NIST) has defined another metric 
called Decision Cost Function (DCF) in order to compare different systems at an interesting operating point \cite{b51}. 
DCF is designed to evaluate the overall cost on making decision errors of both missed detection and false alarm,
and has served as a criterion in every NIST Speaker Recognition Challenge with some parameter adjustments in its definition \cite{b52}.

\begin{equation}
    DCF = C_{FR}P_{Target}P_{FR} + C_{FA}(1-P_{Target})P_{FA}\label{eq11}\end{equation}

where $C_{FR}$ ($C_{Miss}$ in \cite{b51}) and $P_{FR}$ are cost and probability of missed detection;
$P_{Target}$ is a priori probability of the specified target speaker; 
$C_{FA}$ ($C_{FalseAlarm}$ in \cite{b51})and $P_{FA}$ are cost and probability of spurious detection.

In our experiment, $C_{FR}$, $C_{FA}$ are set as 1, and $P_{Target}$ is set as 0.05.

\subsection{Results}
We conduct several experiments to check the efficacy our proposed framework. 
Firstly, we build the state-of-the-art baseline as shown in Table~\ref{tab1} and then compare the performance
of our baseline system with other existing works using both minDCF and EER metrics.
Secondly, we describe our experimental results on frequency selection by adjusting the frequency range and
training our baseline system repetitively for several times.
Moreover, performance improvement of our proposed multi-stream framework is also illustrated.
Finally, we choose some typically used feature dimensions reported in literatures and repeat our experiments
as done in 40-d feature dimension in order to see the generalized improvement level of our proposed method.

\emph{Our baseline system.}
As shown in Table~\ref{tab4}, we compare our baseline system with I-Vectors \cite{b46, b53}, X-Vectors \cite{b9}, 
VoxCeleb1's approach \cite{b46} and VoxCeleb2's approach \cite{b47}.

The term I-Vectors was coined around 2009/2010, where the ``I'' stood for ``Identity'' \cite{b52}, and became
widely used in speaker verification from then on. 
This approach uses one component to model variability of both speaker and channel,
and extract sole low-dimensional representations of utterances. 
Since it is not a neural network based encoder,
network type and loss function are not applicable in Table~\ref{tab4}. 

\begin{table*}
    \caption{Comparison of Speaker Verification Results on VoxCeleb1-Test.}
    \label{tab4}
    \setlength{\tabcolsep}{3pt}
    \begin{tabular}{|p{75pt}|p{50pt}|p{50pt}|p{50pt}|p{50pt}|p{50pt}|p{50pt}|}
    \hline
    System& 
    Train set& 
    Feature&
    Encoder&
    Dim.&
    minDCF&
    EER (\%)\\
    \hline
    I-Vectors \cite{b46, b53}$^{\mathrm{a}}$& 
    Vox1&
    super vector$^{\mathrm{b}}$& 
    --&
    400&
    0.73&
    8.8\\ 
    X-Vectors \cite{b9}& 
    Vox1&
    24-d MFBE& 
    TDNN&
    512&
    0.393$^{\mathrm{c}}$&
    4.16\\ 
    A. Nagrani et .al \cite{b46}& 
    Vox1&
    spectro- gram& 
    VGG-M CNN&
    1024&
    0.75&
    10.2\\  
    A. Nagrani et .al \cite{b46}& 
    Vox1&
    spectro- gram& 
    VGG-M CNN&
    256&
    0.71&
    7.8\\  
    J. S. Chung et al. \cite{b47}& 
    Vox2&
    spectro- gram& 
    ResNet-34&
    512&
    0.549&
    4.83\\ 
    J. S. Chung et al. \cite{b47}& 
    Vox2&
    spectro- gram& 
    ResNet-50&
    512&
    0.429&
    3.95\\  
    Our baseline& 
    Vox2&
    40-d MFBE& 
    ResNet-34$^{\mathrm{d}}$&
    512&
    \textbf{0.210}&
    \textbf{2.73}\\
    \hline
    \multicolumn{7}{p{400pt}}{$^{\mathrm{a}}$Work is done in \cite{b46} using method of \cite{b53}. 
    $^{\mathrm{b}}$A supervector is composed by stacking the mean vectors from a Gaussian Mixture Model (GMM).   
    $^{\mathrm{c}}$$P_{Target}$ is 0.01.
    $^{\mathrm{d}}$The depth of our encoder is one quarter of original ResNet-34. 
    $^{\mathrm{e}}$AP: Angular Prototype. }
    \end{tabular}
    \end{table*}

In \cite{b9}, concept of X-Vectors was first proposed by using deep neural network to capture speaker characteristics, 
and it quickly became the dominating approach for speaker recognition. 
The neural network used in X-Vectors is also called as Time-Delay Neural Network (TDNN), which is naturally 
a variation of 1-D CNN. Additionally, X-Vectors is usually post-processed by PLDA back-end for satisfying results.

In \cite{b46}, VoxCeleb1's approach based on VGG-M CNN is proposed. Different from X-Vectors and our baseline system, 
spectrograms is used as feature, and contrastive loss is design as loss function.

In \cite{b47}, VoxCeleb2's approach based on ResNet is proposed by using VoxCeleb2 as train set. 
Similar to our baseline system, speaker verification is treated as a case of metric learning, 
therefore Euclidean distance is applied as criteria of similarity.
Both ResNet-34 and ResNet-50 are studied, and the performance of ResNet-50 is better than ResNet-34.

Details of our baseline system is depicted in Section~\ref{sec:proposedmethod}.
One notable thing is that almost all reference works have no description on frequency range, 
as they might all use full frequency range.
This is common knowledge, which is also the default setting of our baseline system.
However, this common knowledge might also block us from discovering more useful and interesting things,
and that's why we explore the technique of frequency selection and multi-stream.

\begin{figure}
    \begin{center}    
    \includegraphics[width=1.0\textwidth]{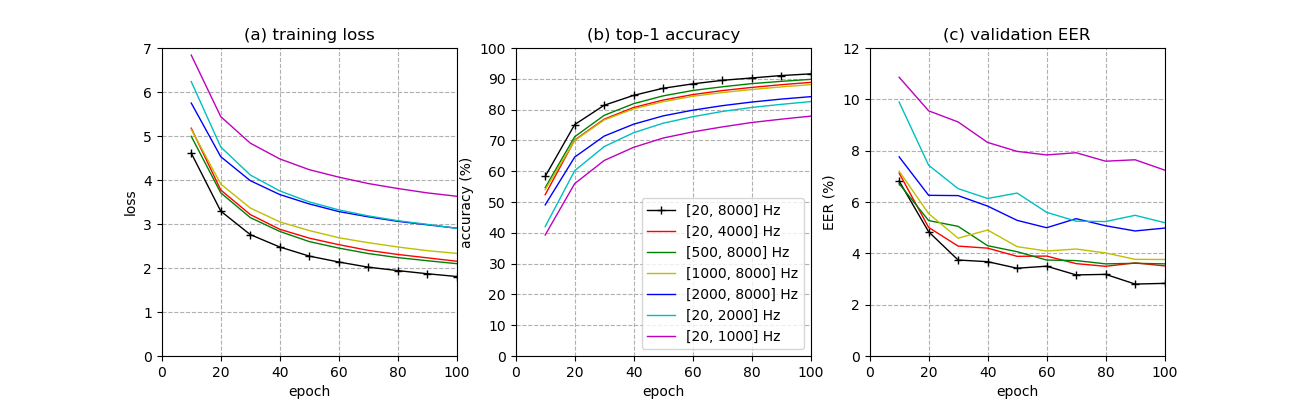}    
    \end{center}    
    \caption{Learning Curves.}    
    \label{fig5}    
    \end{figure}

\emph{Frequency selection and multi-stream.} 
The learning curves of our baseline encoder and other six encoders within different frequency range are illustrated in 
Figure~\ref{fig5}. The curves contain three plots for each encoder: (a)training loss (the smaller, the better), 
(b)top-1 accuracy (the higher, the better), and (c)validation EER (the smaller, the better).

We design three sub-bands for both LF and HF sub-bands, which are LF1 ([20, 1000] Hz), LF2 ([20, 2000] Hz), LF3 ([20, 4000] Hz),
HF1 ([2000, 8000] Hz), HF2 ([1000, 8000] Hz) and HF3 ([500, 8000] Hz) respectively.
It can be perceived from the curves that the baseline encoder within full frequency range has the best performance,
whereas LF1 encoder has the worst performance.
What's more, we can further conclude that the wider the frequency range, the better the system performance 
will be for single-stream system.

The evaluation results of single-stream with different frequency range and multi-stream 
with different combinations of single streams are depicted in Table~\ref{tab5}.
It can be perceived that the combination of FB-stream, LF2-stream and HF2-stream has the best performance.
In addition, it is quite interesting that the combination of top-3 (FB-stream, LF3-stream, HF3-stream) is not the 
best choice, and the reason behind this phenomena might be that only proper frequency range 
instead of wider frequency range can offer relatively the most benefits.
Moreover, LF sub-band and HF sub-band, which is initially designed in Figure~\ref{fig3}, 
is better to be adjusted from non-overlapping to overlapping.
We further conduct more experiments under different dimensions of input feature based on adjusted
frequency segmentation, where LF and HF sub-band are designed to be [20, 2000] Hz and [1000, 8000] Hz respectively.

\begin{table}
    \caption{Evaluation Results of Single-Stream and Multiple-Stream}
    \label{tab5}
    \setlength{\tabcolsep}{3pt}
    \begin{tabular}{|p{100pt}|p{70pt}|p{50pt}|p{50pt}|p{80pt}|}
    \hline
    Stream info.& 
    Frequency range (Hz)&
    minDCF&
    EER (\%)&
    Optimal weight$^{\mathrm{a}}$\\
    \hline
    FB& 
    [20, 8000]& 
    0.2095&
    2.7274&
    --\\
    LF1& 
    [20, 1000]& 
    0.4784&
    6.9301&
    --\\
    LF2& 
    [20, 2000]& 
    0.3622&
    4.9463&
    --\\
    LF3& 
    [20, 4000]& 
    0.2511&
    3.3454&
    --\\
    HF1& 
    [2000, 8000]& 
    0.3420&
    4.7920&
    --\\
    HF2& 
    [1000, 8000]& 
    0.2699&
    3.6326&
    --\\
    HF3& 
    [500, 8000]& 
    0.2609&
    3.4677&
    --\\
    FB + LF1 + HF2& 
    [20, 8000]& 
    0.1694&
    2.331&
    [0.37, 0.35, 0.28]\\
    FB + LF2 + HF1& 
    [20, 8000]& 
    0.1667&
    2.356&
    [0.40, 0.40, 0.20]\\
    FB + LF2 + HF2& 
    [20, 8000]& 
    \textbf{0.1665}&
    \textbf{2.297}&
    [0.39, 0.35, 0.26]\\
    FB + LF2 + HF3& 
    [20, 8000]& 
    0.1717&
    2.339&
    [0.39, 0.35, 0.26]\\
    FB + LF3 + HF2& 
    [20, 8000]& 
    0.1695&
    2.313&
    [0.37, 0.33, 0.30]\\
    FB + LF3 + HF3& 
    [20, 8000]& 
    0.1737&
    2.383&
    [0.48, 0.26, 0.26]\\
    \hline
    \multicolumn{5}{p{350pt}}{$^{\mathrm{a}}$From the left to the right, the weights are for 
    FB-stream, LF-stream, and HF-stream, respectively.}
    \end{tabular}
    \end{table}   

\begin{figure}
    \begin{center}    
    \includegraphics[width=0.55\textwidth]{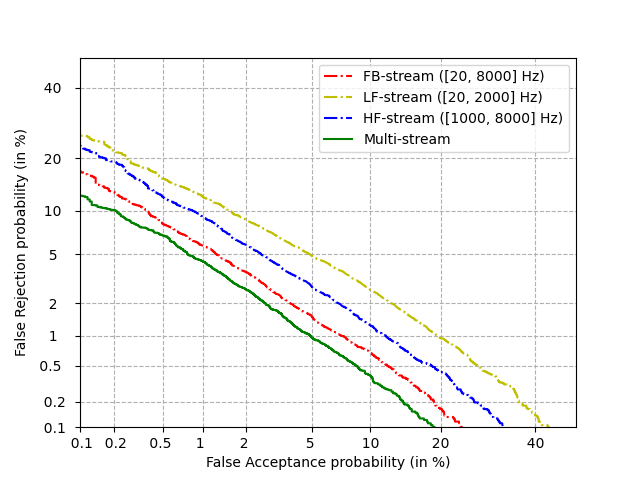}    
    \end{center}    
    \caption{Curves of Detection Error Tradeoff.}    
    \label{fig6}    
    \end{figure}

The Detection Error Tradeoff (DET) curves of FB-stream, LF-stream([20, 2000] Hz), HF-stream ([1000, 8000] Hz)
and Multi-stream are shown with red dash line, yellow dash line, blue dash line and green solid line respectively
in Figure~\ref{fig6}. It can be perceived that our proposed multi-stream framework has comprehensive improvement
compared with our baseline system.

\emph{Improvement vs. Feature Dimension} 
We then conduct experiments by regulating the feature dimension to 32-d and 80-d, 
looking forward to check the efficacy of our proposed method in other dimensions.
The experimental results are illustrated in Table~\ref{tab6}, 
apart from feature dimensions and batch size (the larger the feature dimension, the more memory requirement),
all other configuration are the same for both training and testing.
The batch size for training under feature dimension with 32-d and 80-d are 480 and 200 respectively.

\begin{table}
    \caption{Evaluation Results on Different Feature Dimensions}
    \label{tab6}
    \setlength{\tabcolsep}{3pt}
    \begin{tabular}{|p{60pt}|p{80pt}|p{50pt}|p{50pt}|p{80pt}|}
    \hline
    Feature dim.&
    Stream info.$^{\mathrm{a}}$&
    minDCF&
    EER (\%)&
    Optimal weight$^{\mathrm{b}}$\\
    \hline
    32-d&
    FB-stream&
    0.2316&
    3.077&
    --\\
    &
    LF-stream& 
    0.4108&
    5.346&
    --\\
    &
    HF-stream&
    0.2767&
    3.899&
    --\\
    &
    Multi-stream& 
    0.1805&
    2.489&
    [0.36, 0.31, 0.33]\\
    \hline
    40-d&
    FB-stream&
    0.2095&
    2.727&
    --\\
    &
    LF-stream& 
    0.3622&
    4.946&
    --\\
    &
    HF-stream&
    0.2699&
    3.633&
    --\\
    &
    Multi-stream& 
    0.1665&
    2.297&
    [0.39, 0.35, 0.26]\\ 
    \hline
    80-d&
    FB-stream&
    0.1780&
    2.297&
    --\\
    &
    LF-stream& 
    0.3272&
    4.438&
    --\\
    &
    HF-stream&
    0.2380&
    3.090&
    --\\
    &
    Multi-stream& 
    0.1367&
    1.946&
    [0.36, 0.32, 0.32]\\
    \hline
    \multicolumn{5}{p{350pt}}{$^{\mathrm{a}}$FB-stream: [20, 8000] Hz, LF-stream:  [20, 2000] Hz, HF-stream:  [1000, 8000] Hz. 
    $^{\mathrm{b}}$From the left to the right, the weights are for 
    FB-stream, LF-stream, and HF-stream, respectively.}
    \end{tabular}
    \end{table}  

It can be found from Table~\ref{tab6} that there are significant improvement in both minDCF and EER metrics
from low-dimensional to high-dimensional feature.

\begin{figure}
    \begin{center}    
    \includegraphics[width=0.5\textwidth]{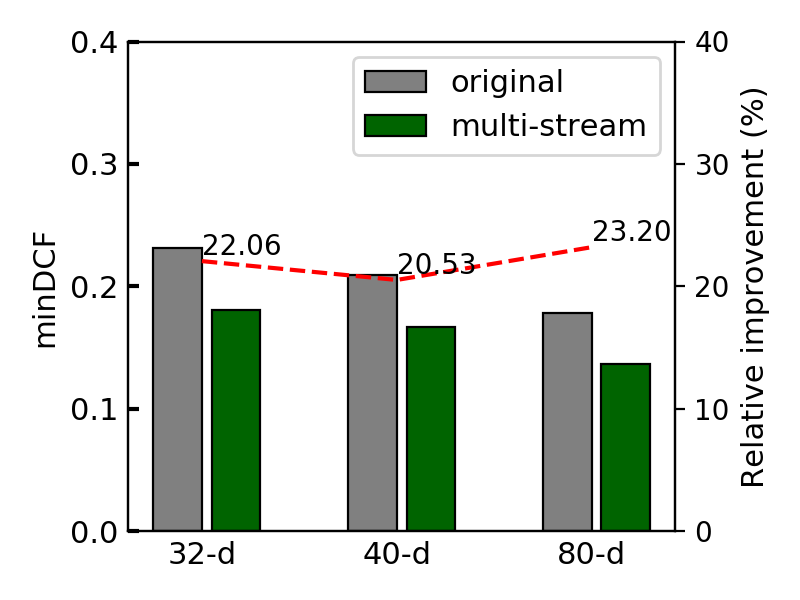}    
    \end{center}    
    \caption{Relative Improvement of minDCF.}    
    \label{fig7}    
    \end{figure}

\begin{figure}
    \begin{center}    
    \includegraphics[width=0.5\textwidth]{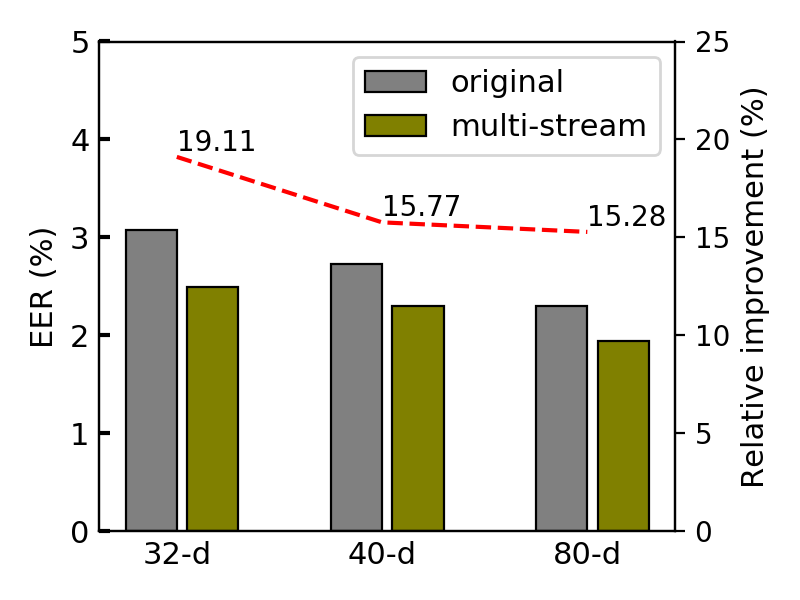}    
    \end{center}    
    \caption{Relative Improvement of EER.}    
    \label{fig8}    
    \end{figure}

The quantitative comparisons are illustrated in Figure~\ref{fig7} and Figure~\ref{fig8}.
Figure~\ref{fig7} shows the relative improvement of minDCF metric, where the left y-axis is 
evaluation result of minDCF with different feature dimensions, and the right y-axis is the 
percentage of relative improvement of minDCF. 
The ``original'' means single-stream with full frequency range, and it is displayed with ``gray'' color.
The ``multi-stream'' means our proposed multi-stream with full frequency selection, 
and it is displayed with ``dark green'' color.
The red dash line illustrates the percentage of relative improvement, and the value are 
22.63 \%, 20.53 \%, and 23.20 \% for 32-d, 40-d and 80-d, respectively.

Figure~\ref{fig8} shows the relative improvement of EER metric, where the left y-axis is 
evaluation result of EER with different feature dimensions, and the right y-axis is the 
percentage of relative improvement of EER. 
The ``original'' and ``multi-stream'' are displayed with ``gray'' and ``olive'' color respectively.
The red dash line illustrates the percentage of relative improvement, and the value are 
19.11 \%, 15.77 \%, and 15.28 \% for 32-d, 40-d and 80-d, respectively.

One remarkable thing is that the relative improvement of EER is not as large as minDCF because the 
objective of our proposed optimal weight search is based on minDCF.
However, this algorithm can be applied to optimal weight search based on EER with few changes.

You might be thinking this is somewhat similar to ensemble scenario, and wonder if 
it can outperform fused system with different training method, such as initialization method?
To answer this question, we conduct additional experiments to make comparison between our proposed framework 
and ensemble of three full-bandwidth models with different initialization methods.

We select three different initialization methods for comparison, which are Kaiming method \cite{b54}, 
Xavier method \cite{b55} and normal distribution.
The experimental results of minDCF and EER under different training epochs are illustrated in Table~\ref{tab7} 
and Table~\ref{tab8} respectively.
The term ``SS'' denotes as Single Stream, whereas ``MS'' denotes as Multiple Stream.
For fair comparison, we only change the initialization method of three full-bandwidth models, and all other training 
configurations remain the same.
All five single stream system are trained for 500 epochs and evaluated every 100 epochs.

\begin{table}
    \caption{Experimental Results of minDCF under Different Training Epochs}
    \label{tab7}
    \setlength{\tabcolsep}{3pt}
    \begin{tabular}{|p{120pt}|p{40pt}|p{40pt}|p{40pt}|p{40pt}|p{40pt}|}
    \hline
    System& 
    100 epochs& 
    200 epochs & 
    300 epochs & 
    400 epochs & 
    500 epochs\\
    \hline
    SS-1: FB-Kaiming-Init$^{\mathrm{a}}$&
    0.2028&
    0.1628&
    0.1688&
    0.1597&
    0.1606\\
    SS-2: LF-Kaiming-Init&
    0.3620&
    0.3223&
    0.3256&
    0.3168&
    0.3125\\
    SS-3: HF-Kaiming-Init&
    0.2574&
    0.2394&
    0.2237&
    0.2299&
    0.2195\\
    SS-4: FB-Xavier-Init&
    0.2323&
    0.2144&
    0.1937&
    0.1825&
    0.1816\\
    SS-5: FB-Normal-Init&
    0.3369&
    0.2933&
    0.2692&
    0.2611&
    0.2544\\
    \hline
    MS-1: SS-1,SS-4,SS-5&
    0.2193&
    0.1888&
    0.1819&
    0.1752&
    0.1722\\
    \textbf{Ours}: SS-1,SS-2,SS-3&
    0.1609&
    0.1333&
    0.1381&
    0.1336&
    0.1310\\    
    \hline
    \multicolumn{6}{p{350pt}}{$^{\mathrm{a}}$Baseline.}
    \end{tabular}
    \end{table}   

\begin{table}
    \caption{Experimental Results of EER with Different Training Epochs}
    \label{tab8}
    \setlength{\tabcolsep}{3pt}
    \begin{tabular}{|p{120pt}|p{40pt}|p{40pt}|p{40pt}|p{40pt}|p{40pt}|}
    \hline
    System& 
    100 epochs & 
    200 epochs & 
    300 epochs & 
    400 epochs & 
    500 epochs\\
    \hline
    SS-1: FB-Kaiming-Init$^{\mathrm{a}}$&
    2.792&
    2.350&
    2.244&
    2.239&
    2.132\\
    SS-2: LF-Kaiming-Init&
    5.109&
    4.212&
    4.133&
    4.143&
    4.042\\
    SS-3: HF-Kaiming-Init&
    3.691&
    3.085&
    3.021&
    2.872&
    2.802\\
    SS-4: FB-Xavier-Init&
    2.840&
    2.553&
    2.239&
    2.096&
    2.106\\
    SS-5: FB-Normal-Init&
    4.578&
    3.844&
    3.649&
    3.376&
    3.260\\
    \hline
    MS-1: SS-1,SS-4,SS-5&
    2.813&
    2.420&
    2.256&
    2.175&
    2.157\\
    \textbf{Ours}: SS-1,SS-2,SS-3&
    2.303&
    1.919&
    1.866&
    1.834&
    1.781\\    
    \hline
    \multicolumn{6}{p{220pt}}{$^{\mathrm{a}}$Baseline.}
    \end{tabular}
    \end{table}   

The system ``SS-1: FB-Kaiming-Init'' is defined as baseline, 
``SS-2'' and ``SS-3'' are both initialized with Kaiming method and trained with partial bandwidth.
``SS-4'' and ``SS-5'' are full-bandwidth models with different initialization method.
The system ``MS-1'' is the ensemble of ``SS-1'', ``SS-4'', and ``SS-5'', which is set to be the comparison object.

\begin{figure}
    \begin{center}    
    \includegraphics[width=0.5\textwidth]{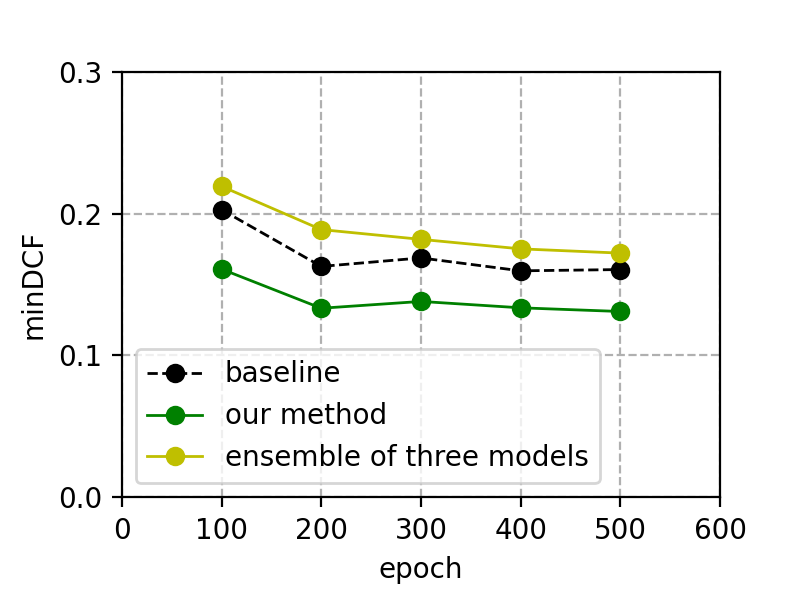}    
    \end{center}    
    \caption{Comparison of minDCF}    
    \label{fig9}    
    \end{figure}

\begin{figure}
    \begin{center}    
    \includegraphics[width=0.5\textwidth]{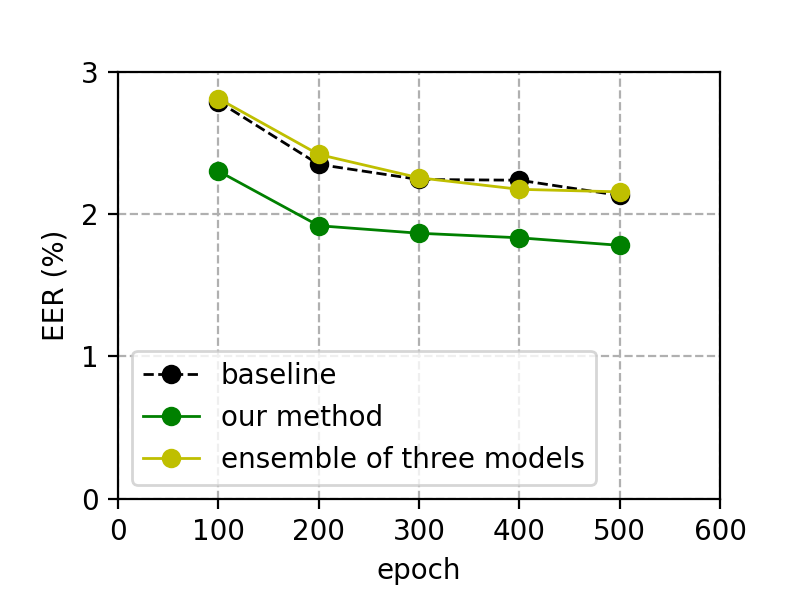}    
    \end{center}    
    \caption{Comparison of EER}    
    \label{fig10}    
    \end{figure}

\begin{figure}
    \begin{center}    
    \includegraphics[width=0.5\textwidth]{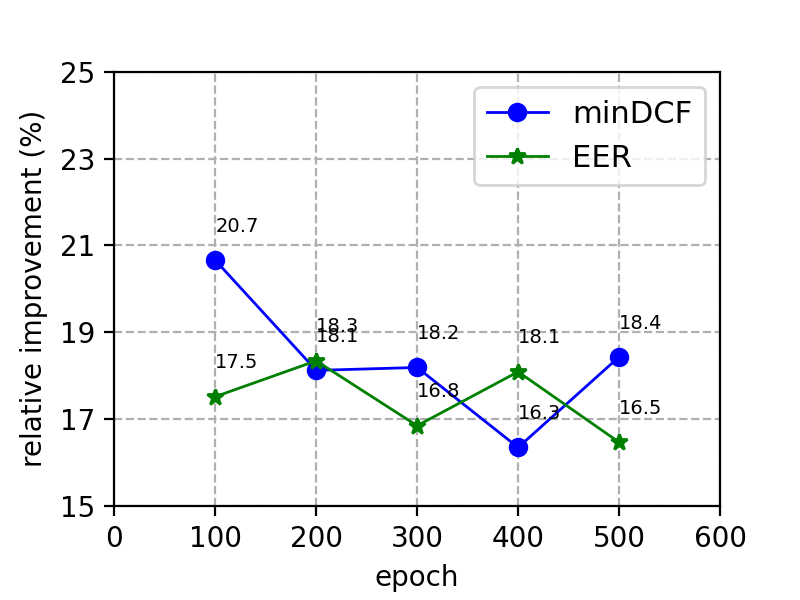}    
    \end{center}    
    \caption{Relative Improvement}    
    \label{fig11}    
    \end{figure}

It can be perceived from the table that our proposed method has significant improvement 
over the whole training period, whereas the ensemble version has almost no improvement but slight degradation.
The graphic comparison can be obtained in Figure~\ref{fig9} and Figure~\ref{fig10}.
Since ``MS-1'' has no improvement, we only plot the relative improvement of our proposed method as shown in 
Figure~\ref{fig11}.

The experimental results show that the ensemble of three full-bandwidth models has poorer performance 
compared to single-stream baseline, whereas our proposed framework has at least 16.3 \% and 16.5 \% 
relative reduction in minDCF and EER respectively.

\section{Future Work}
\label{sec:futurework}
It is obvious that applying frequency selection into speaker recognition with multi-stream can 
improve the performance. But the disadvantage is also obvious due to longer training time and 
larger network size.

As follow-up, can we come up with the trade-off design between size and performance after 
investigating the probability of reducing size meanwhile yielding competitive benefits?
More generally, is frequency selection also effective using other acoustic features such as MFCC, 
and other feature-extracting related techniques such as 
amplitude modulation - frequency modulation (AM-FM) technique \cite{b56}, 
and other neural networks such as Recurrent Neural Network (RNN)?
Furthermore, since MFBE feature is obtained on human perception experiments, 
is it possible to find better features based on machine perception experiments?
These are some remaining questions for us to explore and answer in the future.

\section{Conclusion}
\label{sec:conclusion}
In this paper, we propose a novel neural network framework based on frequency selection, namely multi-stream CNN,
for robust speaker verification. The reason behind this proposal is that the diversity in temporal embeddings 
across multiple streams, where each stream process ``partial'' features extracted within selected frequency range in parallel, 
could enhance the robustness of acoustic modeling and hence improve the overall performance. 
Moreover, in order to address the problem of massive computation and memory requirements,
we propose a more practical two-stage training method consisting of 
``Stage 1: Sequential training of each stream'' and ``Stage 2: Searching for optimal fusion weight''.

To validate our proposed method, we conduct various training and testing experiments using PyTorch library 
based on VoxCeleb dataset, and make comprehensive comparison on the experimental results.
The experimental results demonstrate the efficacy of our proposed method.
The technique derived in our paper can be treated as a variant of metric learning for speaker verification.

\bio{}Wei, Yao\\ works as an associate professor with College of Electric Engineering, 
Zhejiang University of Water Resources and Electric Power.  
He received the M.S degree in Control Theory and Control Engineering from Zhejiang University of Technology 
and the Ph.D. degree in Electric Engineering from Zhejiang University, in 2002 and 2015 respectively. 
His research interests include speech recognition, image processing, embedded system, 
power electronic, etc.

\bio{}Shen, Chen\\ works for Delta Electronics (China) as a senior engineer.
He is currently a software supervisor of one energy solution team, and holds three patents. 
His research interests include internet of things, energy management system,
speech enhancement, speaker recognition, etc.

\bio{}Jiamin, Cui\\ works as a teacher of the College of Electrical engineering, 
Zhejiang University of Water Resources and Electric Power. 
His research interests include communication power supply, intelligent battery charger, 
machine learning in power electronics, etc.

\bio{}Yaolin, Lou\\ works for the College of Electrical Engineering, 
Zhejiang University of Water Resources and Electric Power in 2019. 
His research interest are New Energy Science and Engineering, 
wind power controller system design and simulation, and the electric power system with renewable energy.

\label{lastpage}
\end{document}